\documentclass[prb,twocolumn,superscriptaddress]{revtex4-2}

\usepackage{natbib} 
\usepackage[dvipdfmx]{graphicx} 
\usepackage{dcolumn}
\usepackage{bm}
\usepackage{color}
\usepackage{ulem}
\usepackage{lineno}

\graphicspath{{fig/}}
\usepackage{hyperref}

\begin{document}

\preprint{APS/123-QED}

\title{
Theoretical study on ambient pressure superconductivity in La$_3$Ni$_2$O$_7$ thin films : structural analysis, model construction, and robustness of $s\pm$-wave pairing 
} 

\author{Kensei Ushio}
\affiliation{Faculty of Engineering, Tottori University, 4-101 Koyama-cho Minami, Tottori, Tottori 680-8552, Japan}
\author{Shu Kamiyama}
\affiliation{Department of Physics, University of Osaka, 1-1 Machikaneyama-cho, Toyonaka, Osaka 560-0043, Japan}
\author{Yuto Hoshi}
\affiliation{Faculty of Engineering, Tottori University, 4-101 Koyama-cho Minami, Tottori, Tottori 680-8552, Japan}
\author{Ryota Mizuno}
\affiliation{Forefront Research Center, University of Osaka, 1-1 Machikaneyama-cho, Toyonaka, Osaka 560-0043, Japan}
\author{Masayuki Ochi}
\email{ochi@presto.phys.sci.osaka-u.ac.jp}
\affiliation{Department of Physics, University of Osaka, 1-1 Machikaneyama-cho, Toyonaka, Osaka 560-0043, Japan}
\affiliation{Forefront Research Center, University of Osaka, 1-1 Machikaneyama-cho, Toyonaka, Osaka 560-0043, Japan}
\author{Kazuhiko Kuroki}
\email{kuroki@presto.phys.sci.osaka-u.ac.jp }
\affiliation{Department of Physics, University of Osaka, 1-1 Machikaneyama-cho, Toyonaka, Osaka 560-0043, Japan}
\author{Hirofumi Sakakibara}
\email{sakakibara@tottori-u.ac.jp}
\affiliation{Faculty of Engineering, Tottori University, 4-101 Koyama-cho Minami, Tottori, Tottori 680-8552, Japan}
\affiliation{Advanced Mechanical and Electronic System Research Center(AMES), Tottori University, 4-101 Koyama-cho Minami, Tottori, Tottori 680-8552, Japan}

\date{\today}

\begin{abstract}
We theoretically study ambient pressure superconductivity in thin films of La$_3$Ni$_2$O$_7$. We construct model Hamiltonians adopting the crystal structure theoretically determined by fixing the in-plane lattice constant to those substrates examined in the experiment. We also construct a model based on the experimentally determined lattice structure.  To the models obtained, we apply the fluctuation exchange approximation, which takes into account the full momentum and frequency dependencies of the Green function and the pairing interaction. We find that the electronic structure, including the presence/absence of the so-called $\gamma$-pocket (the Fermi surface originating from the top of the $d_{3z^2-r^2}$ bonding band) depends on the crystal structure adopted and/or the presence/absence of $+U$ correction in the band structure calculation. Nonetheless, $s\pm$-wave pairing symmetry remains robust regardless of these details in the band structure. The robustness of the $s\pm$-wave pairing mainly owes to the fact that it is mediated by finite energy spin fluctuations, which are insensitive to the details of the Fermi surface topology and give rise to a nearly-momentum-independent gap function for the interlayer $d_{3z^2-r^2}$ pairing in the orbital representation. On the other hand, $T_c$ being halved from that of the pressurized bulk can only be understood by adopting the model with small $|t_\perp|$ derived from the experimentally determined crystal structure, at least within the present FLEX approach, although there may remain some other possibilities beyond this approach for the origin of the reduced $T_c$.
\end{abstract}

\pacs{74.20.Mn,74.70.−b}
\maketitle
\section{Introduction}

The discovery of high temperature superconductivity with a $T_c$ near 80 K in a bilayer nickelate La$_3$Ni$_2$O$_7$ has kicked-off a new avenue in the field of condensed matter physics~\cite{MWang,MWangReview}.  Although there has been a debate regarding the bulkiness of the superconductivity, partially substituting La by Pr has resulted in a significant enhancement of the superconducting volume fraction~\cite{LaPr327}. Back in 2017, we theoretically suggested that La$_3$Ni$_2$O$_7$ may be a good starting point for realizing a nearly half-filled bilayer Hubbard model with large interlayer hopping $t_\perp$~\cite{Nakata}, which was shown to exhibit high $T_c$ $s\pm$-wave superconductivity~\cite{KA,MaierScalapino,Nomurabilayer}, but since the material was known to be not superconducting at that time even under pressure~\cite{Hosoya_2008}, we made some speculations regarding the reason for the absence of superconductivity. It now turns out that a key for realizing superconductivity is applying pressure, which induces crystal structures with higher symmetries~\cite{MWang,JACSinsituXRD}. In fact, the strategy for achieving superconductivity by realizing high symmetry crystal structures under pressure has worked also for a trilayer nickelate La$_4$Ni$_3$O$_{10}$, where we theoretically predicted~\cite{sakakibara4310} structural transition from monoclinic to tetragonal symmetries under pressure as well as the occurrence of the superconductivity in the tetragonal phase using the same fluctuation exchange approximation (FLEX), which successfully gave calculation results for $s\pm$-wave superconductivity consistent with $T_c=80$~K in the bilayer La$_3$Ni$_2$O$_7$~\cite{sakakibarala327} (see the Appendix).  Indeed, superconductivity in La$_4$Ni$_3$O$_{10}$ was experimentally realized under pressure in the tetragonal phase~\cite{sakakibara4310,La4Ni3O10Nature,PhysRevX.15.021005,NagataLa4310,La4310Chin}.

Ever since the discovery of high $T_c$ superconductivity in La$_3$Ni$_2$O$_7$, there has been a strong desire to realize superconductivity at ambient pressure in these multilayer nickelates. A reasonable strategy is to seek for materials having a crystal structure with high symmetry at ambient pressure. Theoretically, materials containing actinium~\cite{2403.11713,PhysRevMaterials.8.044801} have been proposed as candidates that can take tetragonal crystal structure at ambient pressure, but they have not been synthesized experimentally, to our knowledge.  We have recently proposed theoretically a new material Sr$_3$Ni$_2$O$_5$Cl$_2$, which we have predicted to take a crystal structure with tetragonal $I4/mmm$ symmetry at ambient pressure~\cite{Sr3252}. This material had never been synthesized before, but right after our proposal, Yamane {\it et al.} succeeded in growing single crystals using high pressure synthesis technique~\cite{Yamane3252}. The structural analysis on the material proved that the crystal structure indeed has $I4/mmm$ symmetry at ambient pressure, although superconductivity is yet to be observed.

Very recently, superconductivity at ambient pressure in bilayer nickelates has finally been realized in thin films of La$_3$Ni$_2$O$_7$~\cite{Ko2025} and (La,Pr)$_3$Ni$_2$O$_7$~\cite{QKXuePrNature,HwangPrNature} on SrLaAlO$_4$ substrate. The observed onset $T_c$ is around 40 K. It is considered that the compressive strain coming from the substrate with a small lattice constant plays the role of applying pressure in bulk systems. In fact, the structural analysis shows that the space group of these thin film systems is $I4/mmm$ (or at least very close to it)~\cite{2501.08204,QKXuePrNature,QKXueThin}.
Structural and electronic analysis on La$_3$Ni$_2$O$_7$ and other related thin films has been performed in a number of theoretical studies~\cite{HirschfeldThin,BotanaThinPRB,2502.01624,2501.08204,85qv-ncxb,DXYao_2503.17223,2501.14665,Rhods_tetra}, some even before the discovery of ambient pressure superconductivity~\cite{HirschfeldThin,BotanaThinPRB}.
In particular, Refs.~\cite{2502.01624,2501.08204,Rhods_tetra,85qv-ncxb}  investigate which structural symmetry is more realistic for thin films on SrLaAlO$_4$ substrates. By performing DFT structural optimization, these studies support the realization of $I4/mmm$ symmetry.
In Ref.~\cite{DXYao_2503.17223}, DFT slab calculation and model construction were performed varying the thickness of the slab. In any case, there appears to be a large deviation between the experimentally determined structure in Ref.~\cite{QKXueThin} and the one obtained from DFT calculations in that the former observes a rather large Ni-Ni distance of 4.28 \AA, resulting in a significantly small $|t_{\perp}|$ close to 0.4 eV~\cite{QKXueThin,Shao9t6n-jqr5,2504.16372}, while the models obtained from DFT structural optimization give $|t_\perp|$ around 0.6 eV~\cite{2502.01624,DXYao_2503.17223}, which does not differ largely from those in models obtained for pressurized bulk~\cite{Luo,sakakibarala327}.

Since superconductivity is now observed at ambient pressure, various probes can be used to investigate the electronic structure that gives rise to the pairing state. In fact, there is already a hot debate going on, where one of the issues is the presence/absence of the so-called $\gamma$ Fermi surface; the Fermi surface around the wave vector $(\pi,\pi)$ that originates from the top of the $d_{3z^2-r^2}$ bonding band. In Ref.~\cite{PengARPES}, an angle resolved photoemission (ARPES) study indicates the presence of the $\gamma$ pocket, while the ARPES study in Ref.~\cite{2504.16372} shows the absence of it.
This is considered to be important because, apart from the thin films, there has been a debate regarding the sensitivity of the pairing symmetry; while some studies indicate sensitivity on the parameter values and/or the Fermi surface structure~\cite{Shao9t6n-jqr5,Eremindwave,LuoDXYao_npjqm,HanghuiChen,PhysRevB.109.104508,PhysRevB.110.024514,Jiang_2024,InoueOnari,2503.18877}, while a random phase approximation study that takes into account the full momentum and frequency dependencies of the Green function and the pairing interaction finds robustness of $s\pm$-wave superconductivity~\cite{Gao2502.19840}. Thanks to the ambient pressure superconductivity, the superconducting gap itself has also been probed experimentally by ARPES~\cite{QKXue2502.17831} and by STM/STS~\cite{HHWen2506.01788}, which seem to be not consistent with pure $d$-wave pairing. Quite recently, pressure effects on superconductivity in thin films has also been examined systematically~\cite{Osada}.

Given the above circumstances, we construct model Hamiltonians adopting the crystal structure theoretically determined by fixing the in-plane lattice constant to those of the three different substrates examined in the experiment~\cite{Ko2025}. We adopt PBEsol functional for structural optimization assuming $Amam$, while we compare PBEsol and PBEsol$+U$ cases in the band structure calculation. We find a trend where the interlayer Ni-O-Ni bond angle approaches, but does not reach, 180 degrees as the in-plane lattice constant becomes smaller, qualitatively consistent with previous theoretical studies~\cite{BotanaThinPRB,2502.01624,85qv-ncxb}. For the case of SrLaAlO$_4$ substrate in particular, we also consider two additional crystal structures that have $I4/mmm$ symmetry; one obtained by DFT structural optimization within $I4/mmm$ symmetry, and the other determined experimentally in Ref.~\cite{QKXueThin}. To the models obtained, we apply FLEX, which takes into account the full momentum and frequency dependencies of the Green function and the pairing interaction. We find that the electronic structure, including the presence/absence of the $\gamma$ pocket, depends on the crystal structure adopted and/or the presence/absence of $+U$ correction in the band structure calculation, but the $s\pm$-wave pairing symmetry remains robust. On the other hand, $T_c$ being halved from that of the pressurized bulk can only be understood by adopting the model with small $|t_\perp|$ derived from the experimentally determined crystal structure, at least within the present FLEX approach, although there may remain some other possibilities beyond this approach for the origin of the reduced $T_c$.

\section{Method}

\begin{figure}
	\includegraphics[width=9cm]{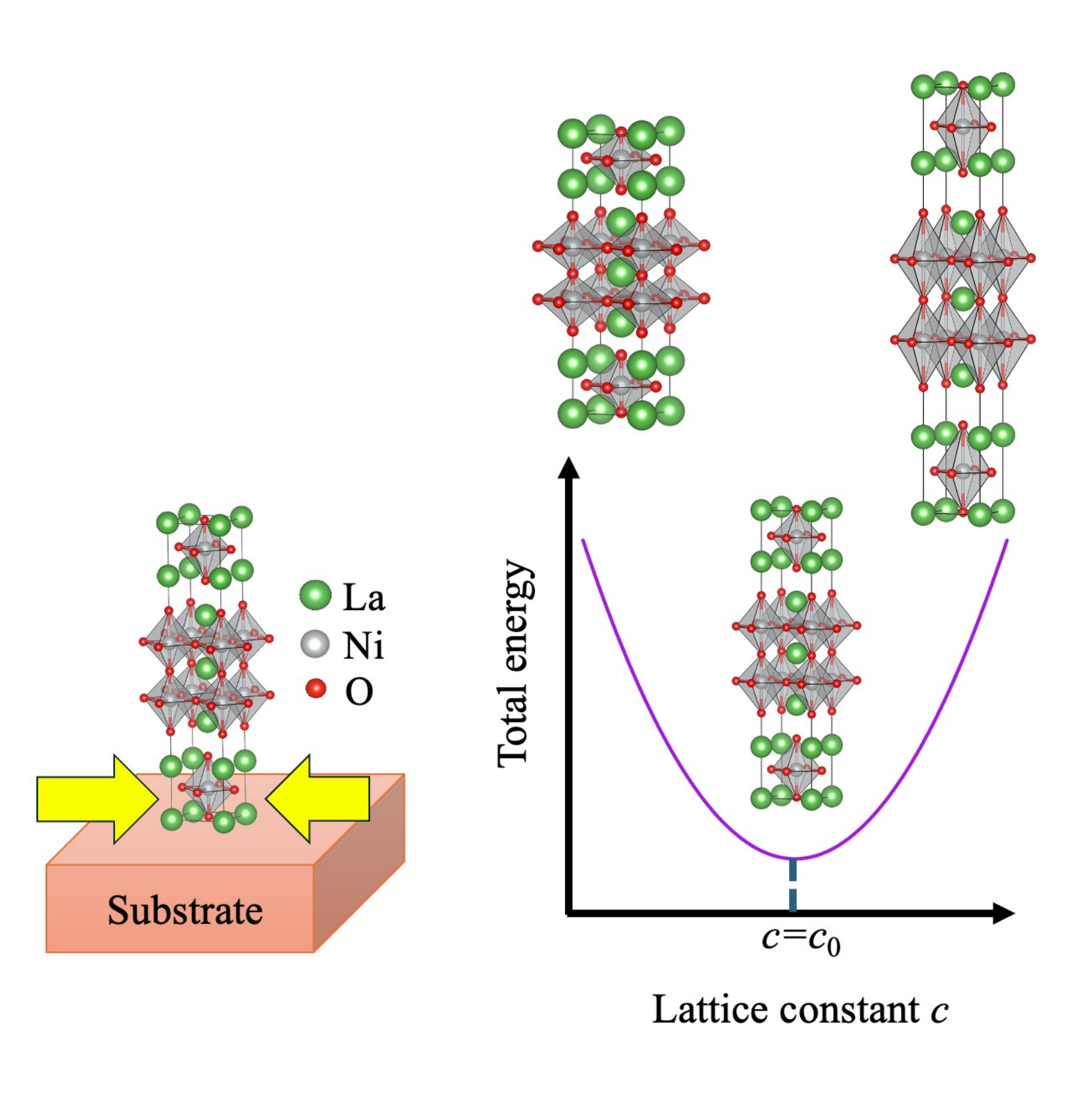}	
	\caption{An image of the determination procedure for structural parameters in this study. 
	Crystal structure is drawn by VESTA~\cite{VESTA}.
	}
	\label{fig1}
\end{figure}

We perform first-principles structural optimization to determine the crystal structure, where we consider a bulk system with the in-plane lattice constants fixed to that of the substrates. We use the PBEsol exchange-correlation functional~\cite{PBEsol} and the projector augmented wave (PAW) method as implemented in Vienna {\it ab initio} Simulation Package (VASP)~\cite{vasp1,vasp2,vasp3,vasp4}.
Core-electron states in PAW potentials were [Kr]4$d^{10}$, [Ar], and [He] for La, Ni and O, respectively. We use a plane-wave cutoff energy of 550 eV for Kohn-Sham orbitals. 
We take an 8 $\times$ 8 $\times$ 8 ${\bm k}$-mesh~\cite{kconv}, and an energy width of 0.1 eV for Gaussian smearing.

We consider three cases of the substrate as examined in Ref.~\cite{Ko2025}: (LaAlO$_3$)$_{0.3}$(Sr$_2$TaAlO$_6$)$_{0.35}$ (LSAT),  LaAlO$_3$(LAO), and SrLaAlO$_4$(SLAO), in which the lattice constants $a$ are 3.868 \AA, 3.787 \AA,  and 3.756 \AA, respectively. 
The lattice mismatches between the experimental value (=3.833 \AA~\cite{Ko2025}) of the bulk system and the substrates are $+0.9$, $-1.1$, and $-2.0$~\% for LSAT, LAO and SLAO cases, respectively.
We determine the theoretical value of the lattice constant $c=c_0$ so that the total energy is the lowest within this restriction. To determine the total energy for given $c$, we perform optimization calculation of atomic positions.
The convergence condition is for the Hellmann-Feynman force to be less than 0.001 eV \AA$^{-1}$ for each atom.
By fitting the plots of the total energy with quadratic function, we obtain the lattice constant $c_0$. A schematic figure of this procedure is shown in Fig.~\ref{fig1}.
The calculations are performed assuming $Amam$ structure, as suggested in the experiment for bulk system at ambient pressure~\cite{MWang}.

Apart from the crystal structure determined by DFT calculation as described above, we also employ two additional structures in the case of SLAO substrate. One is the structure obtained by DFT optimization assuming $I4/mmm$ symmetry, and the other is the experimentally determined structure with $I4/mmm$ symmetry provided in Ref.~\cite{QKXueThin}.

To analyze superconductivity, we construct model Hamiltonians for the crystal structures determined by the above procedure.
For band structure calculations, we include $+U$ correction in the Dudarev's formulation~\cite{GGAU} with $U = 3$ eV.
For comparison, we also performed band calculation with $U=0$. Then we extract maximally localized Wannier functions~\cite{Marzari,Souza} using the {\footnotesize WANNIER90} software~\cite{Wannier90}, 
by which we also obtain the hopping parameters among the Wannier functions. 
We have constructed a two-orbital bilayer tight-binding model 
choosing the Ni-$3d_{x^2-y^2}$ and $3d_{z^2-r^2}$ orbitals as initial projections. 
Although the symmetry of the resulting Wannier functions slightly differs from the initial projections due to the Ni-O octahedral tilting in $Amam$ symmetry, however, we label these Wannier orbitals based on their initial projections for the sake of convenience.
For this purpose, we take a 12 $\times$ 12 $\times$ 12 ${\bm k}$-mesh for a primitive unit cell of $I4/mmm$ structure, while we take an 8$\times$8$\times$8 ${\bm k}$-point mesh for $Amam$, which has four Ni sites per unit cell.
The same $k$-mesh is maintained between the first-principles calculations and the Wannier fitting procedure.
Then we obtained a two-orbital bilayer Hubbard model by adding only the onsite interaction terms to the tight-binding terms.
As for the strength of the interactions, the parameters presented in Ref.~\cite{Werner} are adopted.
Namely, intraorbital repulsion  $U=3.79$ $(3.58)$ eV for the $d_{x^2-y^2}$ ($d_{3z^2-r^2}$) orbital, 
 interorbital repulsions $U'=2.39$ eV, Hund's coupling and the pair hopping terms $J=J'=0.61$ eV are adopted.
We use this consistent set of $U,U', J, J'$ values across the calculations, and the robustness of our conclusions is supported by the fact that an eigenvalue $\lambda$ of the linearized Eliashberg equation, as described in the next paragraph, exhibits a relatively low sensitivity to the exact choice of the values within a physically reasonable range. This insensitivity has been explicitly demonstrated in our previous systematic studies~\cite{sakakibarala327,Sr3252}.

We take into account the electron correlation effects for the two-orbital bilayer Hubbard model within FLEX approximation~\cite{Bickers,Bickers1991} as was done in Ref.~\cite{sakakibarala327}. We calculate the self-energy induced by the spin-fluctuation formulated as shown in the literatures~\cite{Lichtenstein,mFLEX1,mFLEX2}, in a self-consistent calculation.
The formulation of FLEX is given in the APPENDIX.
The real part of the self-energy at the lowest Matsubara frequency is subtracted in the same manner with Ref.~\cite{Ikeda_omega0}
to maintain the band structure around the Fermi level obtained by first-principles calculation.
The obtained Green function and the pairing interaction, mediated mainly by spin fluctuations, are plugged into the linearized Eliashberg equation.
Since the eigenvalue $\lambda$ of the linearized Eliashberg equation reaches unity at $T=T_c$, 
we adopt it as a measure of superconductivity at a fixed temperature, $T=0.01$ eV. 
The eigenfunction of the linearized Eliashberg equation corresponds to the anomalous self energy in the original (non-linearized) Eliashberg equation, but for convenience, we will call the eigenfunction (with the largest eigenvalue)  at the lowest Matsubara frequency $i\omega$(=$i\pi k_{\rm B}T$) the ``superconducting gap function''. 
In this work, the gap functions are calculated in the orbital basis and subsequently transformed into the band basis using the unitary matrix derived from the diagonalization of the unperturbed normal-state Hamiltonian.
We take a 32$\times$32$\times$4(16$\times$16$\times$4) ${\bm k}$-point mesh and 2048 Matsubara frequencies for $I4/mmm$($Amam$).
For all cases of the present calculation, we obtain the eigenfunction of $s\pm$-wave superconductivity.
Note that the pairing interaction kernel in this equation was obtained from the FLEX Green function as a purely electronic one (i.e. phonon-mediated pairing interaction was not considered), which is mainly dominated by spin fluctuations in the present case.

\section{Results}

\subsection{Structural parameters}

\begin{figure}
	\includegraphics[width=9cm]{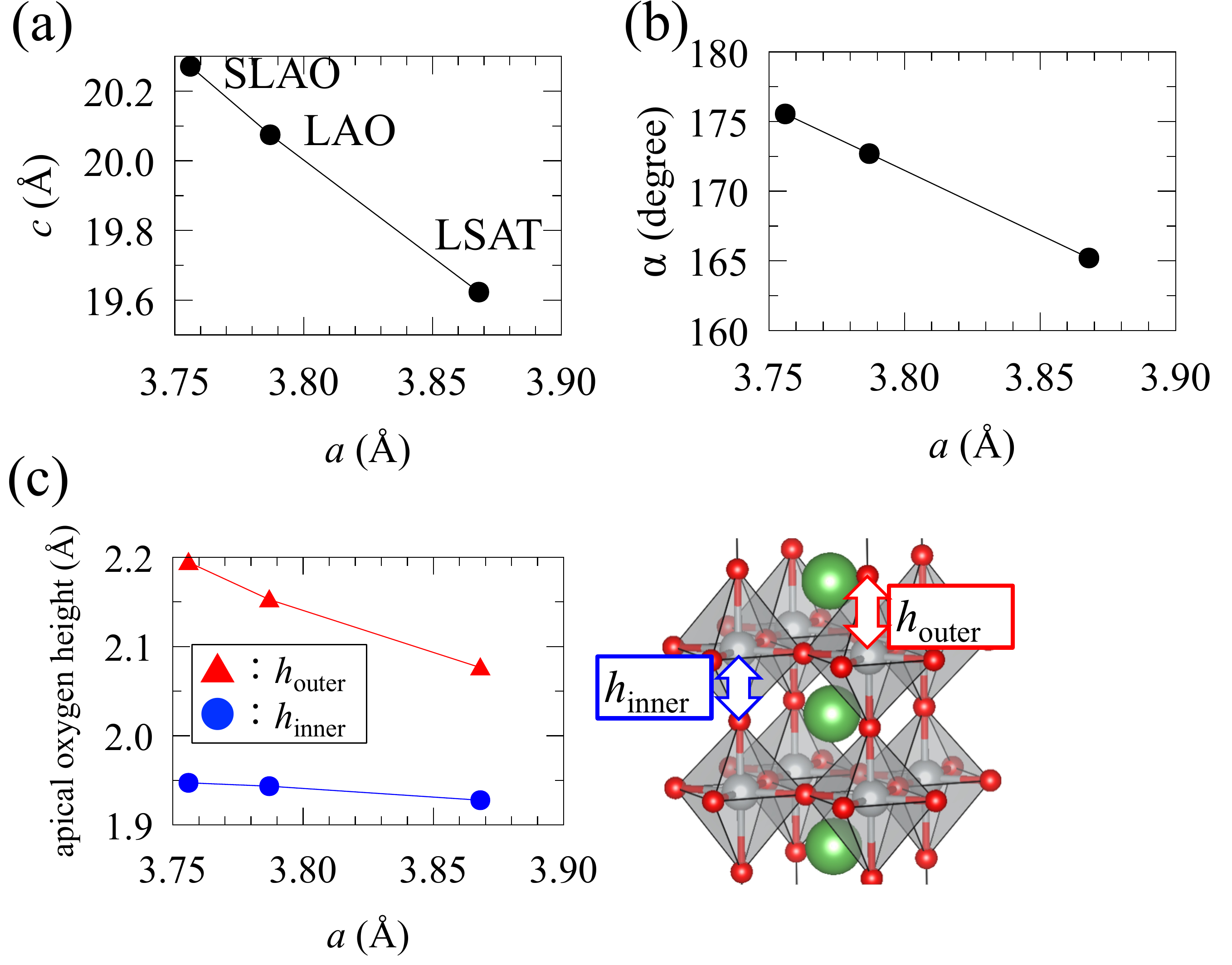}	
	\caption{Structural parameters plotted against the in-plane lattice constant $a$.
	Panels (a), (b), and (c) display the lattice constant $c$, interlayer Ni-O-Ni bond angle $\alpha$, the height of the apical oxygens $h_{\rm outer}$ and $h_{\rm inner}$, respectively.
	The definition of $h_{\rm outer}$ and $h_{\rm inner}$ is shown in the inset.
	}
	\label{fig2}
\end{figure}

In Fig.~\ref{fig2}(a), we plot the lattice constant $c$ as a function of the lattice constant $a$. It is seen that the structure is elongated in the $c$ direction for  smaller $a$.
For the SLAO case, the theoretical value ($c=20.27$\AA) is smaller than the experimental value reported by Yue {\it et al.} ($c=20.819$\AA~\cite{QKXueThin}).
The effect of this difference will be discussed in Sec. III C.

In Fig.~\ref{fig2}(b), we plot the interlayer Ni-O-Ni bond angle $\alpha$ as a measure of the proximity towards the structural transition point from $Amam$ to $I4/mmm$ phase. The bond angle increases toward 180 degrees by reducing the lattice constant $a$, which is qualitatively consistent with a trend captured in a previous theoretical study~\cite{85qv-ncxb}.
Within the present calculation scheme, however, the angle $\alpha$ does not reach 180 degrees even in the SLAO case. Nonetheless, we do not intend to conclude that the structure of the thin film on SLAO substrate has $Amam$ symmetry. In fact, for comparison, we also perform structural optimization assuming $I4/mmm$ symmetry and have found that the energy difference between $Amam$ and $I4/mmm$ phase is about few meV per unit cell, which is small. This suggests the difficulty in judging which phase is more stable between $Amam$ and $I4/mmm$ through first-principles calculations.
As mentioned in the Introduction, experiments suggest realization of $I4/mmm$ symmetry within the experimental error bar~\cite{2501.08204,QKXueThin}.

\subsection{Band structure and hopping integrals for the theoretically determined lattice structure}

\begin{figure}
	\includegraphics[width=9cm]{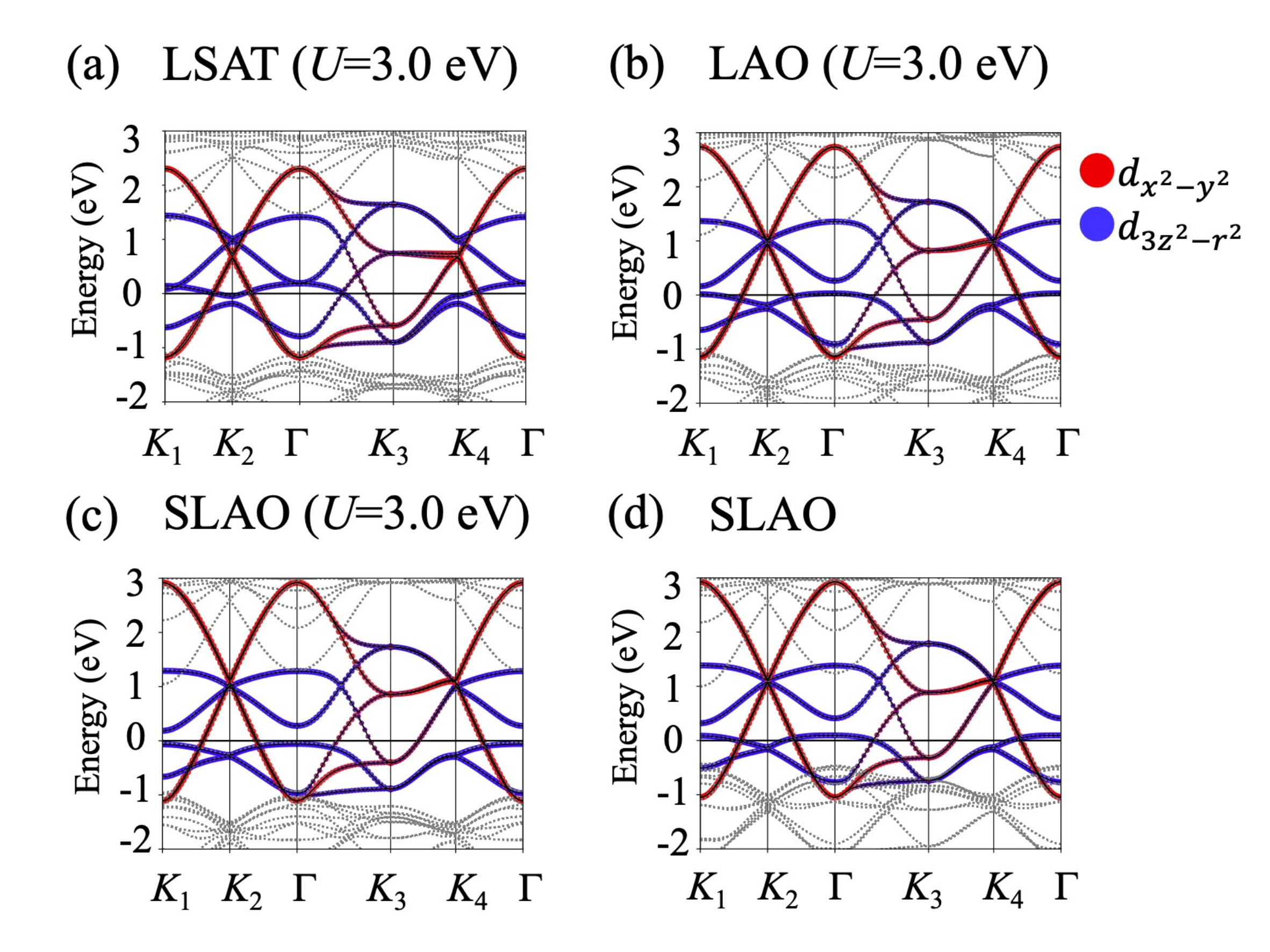}
	\caption{(black solid and gray dashed lines) Energy band dispersion of the two-orbital bilayer models on top of that of first principles calculations, assuming $Amam$ symmetry.
	 Panels (a)-(c) are for LSAT, LAO, and SLAO cases, where the DFT+$U$ method is adopted ($U=3$ eV).
	 Here, the Fermi level is set at $E=0$ eV and
	the weights of the $d_{x^2-y^2}$($d_{3z^2-r^2}$) orbitals are represented by the thickness of the red (blue) line.
	The high-symmetry points are defined as follows: $K_1 = (\mathbf{b}_2 + \mathbf{b}_3)/2$, $K_2 = \mathbf{b}_3/2$, $K_3 = (\mathbf{b}_1 + \mathbf{b}_2)/2$, and $K_4 = \mathbf{b}_2/2$. Here, the reciprocal lattice vectors are given by $\mathbf{b}_1 = (2\pi/a, 0, 0)$, $\mathbf{b}_2 = (0, 2\pi/b, -2\pi/c)$, and $\mathbf{b}_3 = (0, 2\pi/b, 2\pi/c)$, where $a, b$ and $c$ are the lattice constants.
	In panel (d), for comparison, the result of DFT calculation without $+U$ correction for the SLAO case is shown.
	}
	\label{fig3}
\end{figure}

\begin{figure}
	\includegraphics[width=9cm]{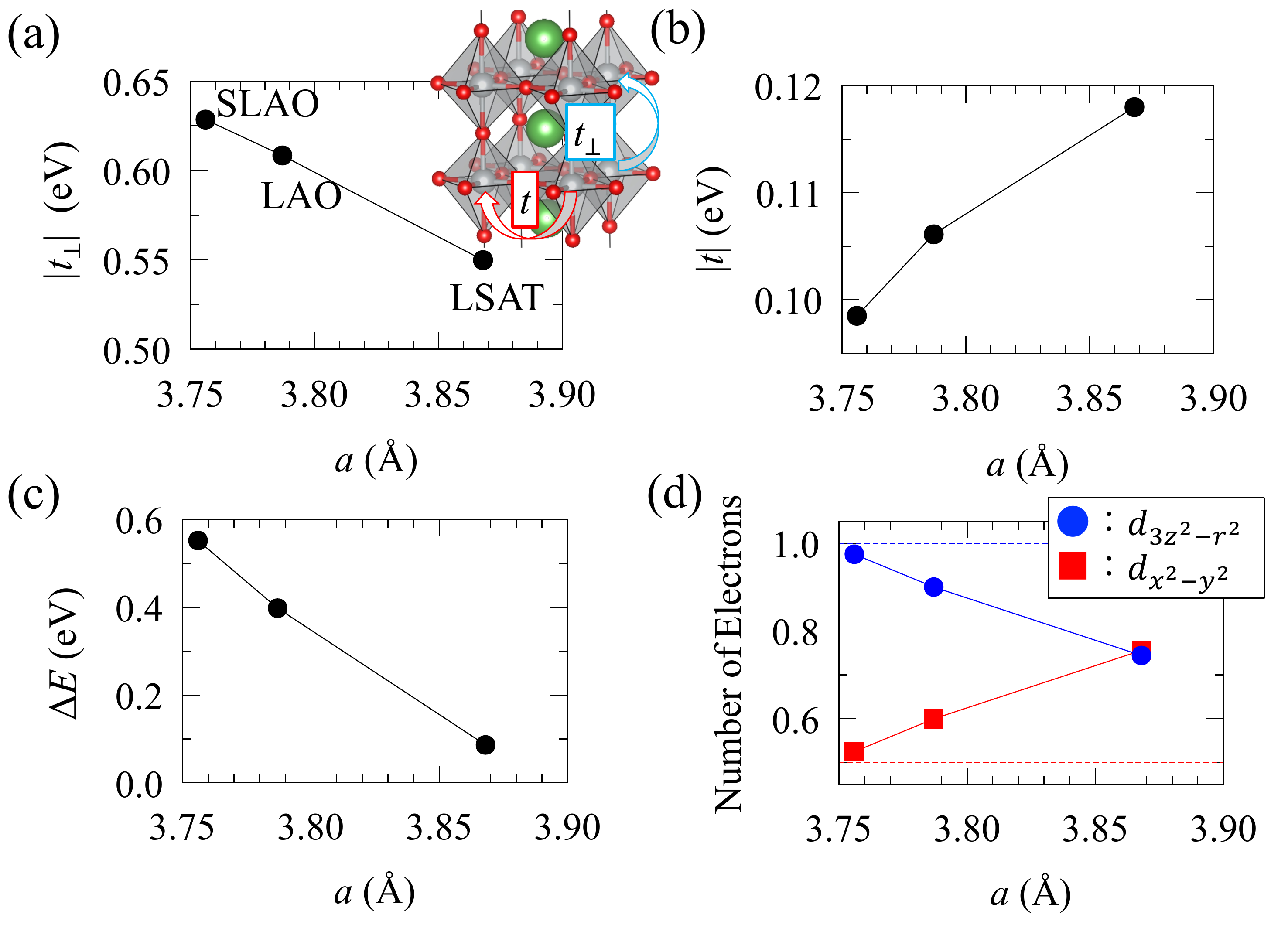}
	\caption{Hopping integrals (a) $t_{\perp}$ and (b) $t$ of the 
	$d_{3z^2-r^2}$ orbitals and (c) the level offset $\Delta E$ are plotted against the lattice constant $a$.
	Here, $\Delta E=E_{x^2-y^2}-E_{3z^2-r^2}$ and $E_{x^2-y^2}$ ($E_{3z^2-r^2}$) are the onsite energies of the $d_{x^2-y^2}$ ($d_{3z^2-r^2}$) Wannier orbital.
	The inset shows the definition of $t_{\perp}$ and $t$.
	Number of electrons per Ni atom are also plotted in panel (d).
	}
	\label{fig4}
\end{figure}

In Fig.~\ref{fig3}, we plot the band structure of the three substrate cases.
The mean position of the $d_{3z^2-r^2}$ bonding band is lower for substrates with smaller in-plane lattice constant $a$. Especially, the top of the bonding band around the $\Gamma$ point (corresponding to the X point in the unfolded Brillouin zone) is above the Fermi level for the LSAT and LAO cases, while it is below for the SLAO case.
This difference can be explained by two tight-binding parameters:
the interlayer hopping integral $t_{\perp}$ of  the $d_{3z^2-r^2}$ orbitals and the level offset 
$\Delta E=E_{x^2-y^2}-E_{3z^2-r^2}$, where $E_{x^2-y^2}$ ($E_{3z^2-r^2}$) is the onsite energy, defined as the onsite and orbital-diagonal component of hopping integral, of the $d_{x^2-y^2}$ ($d_{3z^2-r^2}$) orbital.
As explained in Ref.~\cite{sakakibarala327}, the increase of $|t_\perp|$ and $\Delta E$ results in a downward shift of the bonding band of the $d_{3z^2-r^2}$ orbital.
These parameters increase for smaller lattice constant $a$ as displayed in Fig.~\ref{fig4},
and thus occurs the downward shift of the $d_{3z^2-r^2}$ bonding band relative to the $d_{x^2-y^2}$ bands.
Enhancement of $\Delta E$ for smaller $a$ (Fig.~\ref{fig2}(c) and the inset) can be understood intuitively because higher apical oxygen height $h_{\rm outer}$ enhances crystal field splitting between the $d_{3z^2-r^2}$ and $d_{x^2-y^2}$ orbitals, the former of which is mainly repulsed by the apical oxygen.
On the other hand, the increase of $|t_{\perp}|$ for smaller $a$ seemingly contradicts with the increase of $h_{\rm inner}$.  
In fact, the increase of $|t_{\perp}|$ can be considered as consistent with the increase of the bond angle $\alpha$ because $t_\perp$ is correlated with the strength of the $\sigma$ bonding between the Ni-$d_{3z^2-r^2}$ and the O-$p_z$ orbitals. 
The present results indicate that $t_{\perp}$ is more governed by the bond angle $\alpha$ than the apical oxygen height $h_{\rm inner}$ as far as the dependence on the in-plane lattice constant is concerned.

In Fig.~\ref{fig4}(b), we also plot the value of the in-plane hopping $t$ of the $d_{3z^2-r^2}$ orbitals, in which the deviation from fourfold symmetry is less than $10^{-4}$ eV.
Interestingly, $t$ turns out to be larger for larger $a$. Although the reason for this apparently counter-intuitive behavior is not clear at present, it may be due to the larger tilting of the $d_{3z^2-r^2}$ orbitals, or due to the shorter distance between Ni and the apical oxygens, enhancing the molecular structure of the Wannier orbital. We note that $|t_\perp|$ reaches $\sim 0.6$ eV for SLAO, which is consistent with previous studies that employ theoretically determined crystal structures~\cite{2502.01624,DXYao_2503.17223}. As mentioned in the Introduction, this value of $t_\perp$ is not significantly different from that in the pressurized bulk~\cite{sakakibarala327,Luo}.

Due to the enhancement of the interorbital energy level offset $\Delta E$ for smaller $a$, the number of electrons occupying the $d_{3z^2-r^2}$ ($d_{x^2-y^2}$) orbitals increases (decreases) as $a$ is reduced, as depicted in Fig.~\ref{fig4}(d). For the SLAO substrate in particular, the orbital occupancy of the $d_{3z^2-r^2}$ orbital reaches 0.95 per Ni atom.

\subsection{Superconductivity}

We now turn to the analysis of superconductivity. We consider only the case of the SLAO substrate, and adopt three types of crystal structures; those determined by DFT optimization (i) within $Amam$ or (ii) assuming $I4/mmm$ symmetry and (iii) the one determined experimentally, also having $I4/mmm$ symmetry. For all crystal structures, the band structure calculations are performed using PBEsol+$U$ functional with $U=3$ eV (denoted as theoretical structure ($U=3$ eV) in the following figures) or $U=0$ (denoted as theoretical structure) as noted above, but for case (i) we also study the $U$ dependence within the $+U$ correction scheme.

\begin{figure}
	\includegraphics[width=9cm]{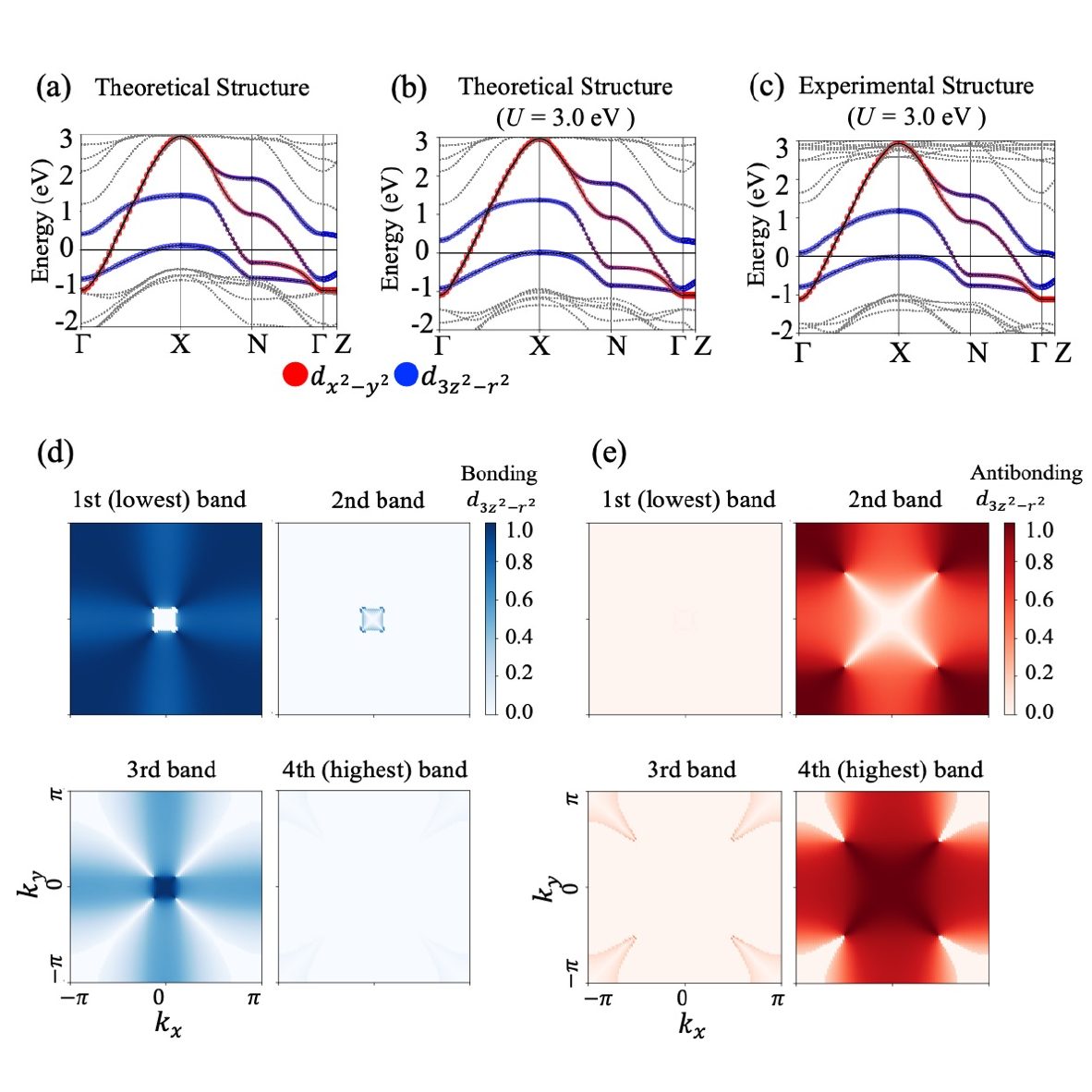}
	\caption{(black solid and gray dashed lines) Energy band dispersion of two-orbital bilayer model on top of that of first principles calculations, assuming $I4/mmm$ symmetry.
	Panels (a) and (b) are for DFT and DFT+$U$($U=3$ eV), respectively. 
	 Here, the Fermi level is set at $E=0$ eV and
	the weights of the $d_{x^2-y^2}$($d_{3z^2-r^2}$) orbitals are represented by the thickness of the red (blue) line.
	In panel (c), the case for DFT+$U$ using experimental structure of Ref.~\cite{QKXueThin} is also shown. For the case of DFT$+U$ ($U=3$eV),  the strength of the $d_{3z^2-r^2}$ (d) bonding and (e) antibonding character, as defined in Eq.(1), is presented for the four bands in the entire Brillouin zone. See the main text for the explanation of the discontinuity in the orbital character.
	}
	\label{fig5}
\end{figure}

\begin{figure*}
	\includegraphics[width=18cm]{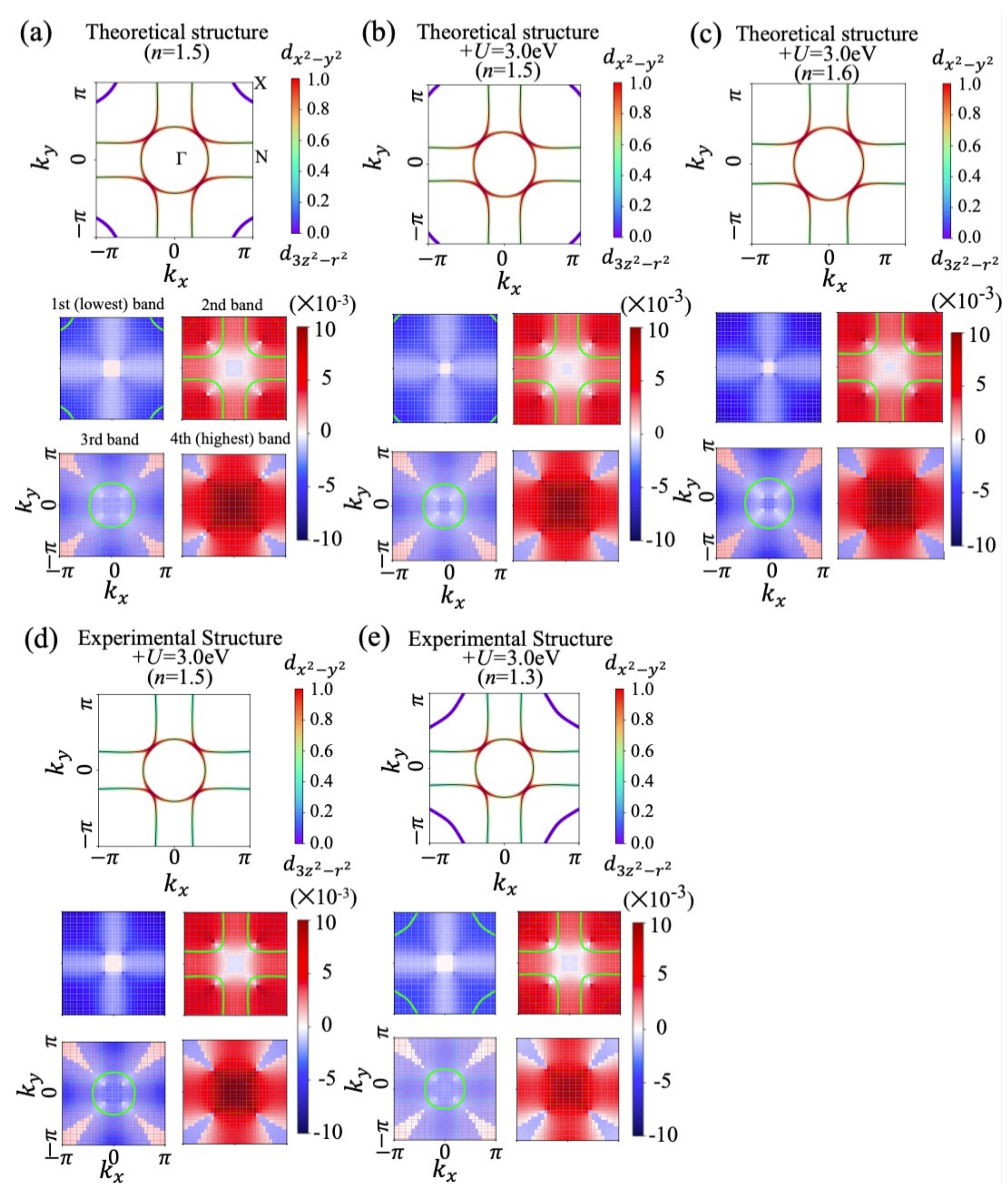}
	\caption{The Fermi surfaces (upper panel) and the gap functions (lower panel) at the $k_z=0$ cross section are presented on the two-dimensional Brillouin zone.
	Panel (a) is for the DFT case and the others are for DFT+$U$($U=3$ eV) cases.
	Fully theoretically derived crystal structure is used in panels (a)-(c)
	while that of the experiment~\cite{QKXueThin} is used in panels (d)(e).
	The band filling (the number of electrons per Ni atom) is displayed in the round brackets.
	For the gap functions, band indices are assigned in ascending order from lower energy as in Fig.~\ref{fig5}(d)(e). 
	The green lines in the lower panels indicate the Fermi surface corresponding to the band indices.
	}
	\label{fig6}
\end{figure*}

\subsubsection{Calculations with I4/mmm symmetry: band structures, Fermi surfaces, and superconducting gap functions}

In Fig.~\ref{fig5}, we plot the band structures for the three $I4/mmm$ cases. For the experimentally determined structure, it can be seen that the $d_{3z^2-r^2}$ bonding-antibonding splitting is small, and both the bonding band top at the X point and the antibonding bottom at the $\Gamma$ point (almost) touches the Fermi level. This is because $|t_{\perp}|$ in this case is as small as $0.448$ eV,  which is only 70 \% of the value derived from theoretically determined structure ($0.636$ eV), consistent with previous theoretical studies~\cite{QKXueThin,Shao9t6n-jqr5,2504.16372} based on the experimentally determined structure.

In the lower panels of Fig.~\ref{fig5}, we depict the strength of the (d) bonding and (e) antibonding $d_{3z^2-r^2}$  character in each band in the entire Brillouin zone for the sake of understanding the gap function in the band representation provided in Fig.~\ref{fig6}. Here, the bonding and antibonding components are defined as the symmetric ($+$) and antisymmetric ($-$) combinations of the $d_{x^2-y^2}$ and $d_{3z^2-r^2}$ Wannier basis, as was introduced in Refs.~\cite{RyeeDMFT, RyeeRPA}.
Hereinafter, we refer to $d_{x^2-y^2}$ and $d_{3z^2-r^2}$ basis as $x,z$ for simplicity.
The transformation between bonding/antibonding and upper/lower layer basis is given using a matrix $A$ as
\begin{eqnarray}
\left( \begin{array}{c}
c_{x(+)}  \\ c_{z(+)} \\ c_{x(-)}  \\ c_{z(-)}
 \end{array}
\right)&=&\frac{1}{\sqrt{2}}\left(\begin{array}{cccc}
1 & 0 & 1 & 0\\
0 & 1 & 0 & 1\\
1 & 0 & -1 & 0\\
0 & 1 & 0 & -1
\end{array}\right)
\left(\begin{array}{c}
c_{x({\rm U})}  \\ c_{z({\rm U})}\\ c_{x({\rm L})}\\ c_{z({\rm L})}\\
\end{array} \right)\nonumber\\
&=&A\left( \begin{array}{c}
c_{x({\rm U})}  \\ c_{z({\rm U})}\\ c_{x({\rm L})}\\ c_{z({\rm L})}\\
\end{array} \right),\label{unimat}
\end{eqnarray}
where $c$ is the electron annihilation operator of $d$ orbitals in upper and lower layers, referred to as U and L, respectively.
In Eq.~(\ref{unimat}), site and spin indices are omitted for simplicity.
The band indices are assigned in ascending order from lower energy, and 
this causes the orbital character to switch between different band indices at $k$-points where band crossings occur. Consequently, apparent discontinuities appear in the orbital character; for instance, the anti-bonding $d_{3z^2-r^2}$ 
character is strong in the fourth (highest) band around $(0,0) \sim (\pi,0)/(0,\pi)$, but this is continued to $\sim (\pi,\pi)$ in the second band, as shown in Fig. \ref{fig5}(a)-(c) and (e).

In the upper panels of Fig.~\ref{fig6}, we plot the Fermi surface for various $I4/mmm$ cases.  For the theoretically determined crystal structure,  at the band filling of $n=1.5$ corresponding to the stoichiometric case, the $\gamma$ pocket around $(\pi,\pi)$ shrinks with the $+U$ correction in the band calculation, and totally vanishes when the band filling is increased to $n=1.6$. For the experimental crystal structure, the top of the $d_{3z^2-r^2}$ bonding band barely touches the Fermi level for $n=1.5$, but when holes are doped, as indicated in the experiment~\cite{PengARPES}, the $\gamma$ pocket appears.

For all of these cases, the gap functions in the band representation are plotted in the lower panels of Fig.~\ref{fig6}. The presence/absence of the $\gamma$ pocket and other details do not largely affect the gap function structure. A common feature is that the gap function exhibits large values at portions of the bands where there is $d_{3z^2-r^2}$ orbital character, as seen by comparing the gap function and Fig.~\ref{fig5}(d)(e). The gap function exhibits positive values at portions of bands 1 and 3 where there is bonding $d_{3z^2-r^2}$ orbital character, while it exhibits negative values in bands 2 and 4 at portions with antibonding $d_{3z^2-r^2}$ orbital character. In fact, the magnitude of the gap function closely resembles that of the strength of the $d_{3z^2-r^2}$ orbital character shown in Fig.~\ref{fig5}(d)(e), indicating that the momentum dependence of the gap function in each band is basically a consequence of the momentum dependence of the orbital weight.

The overall features of the FLEX gap function, which was already presented in our previous study along the high-symmetry path in the Brillouin zone~\cite{sakakibarala327},  well agrees with those obtained in a cellular dynamical mean field theory (cDMFT)~\cite{RyeeDMFT}, which is a non-perturbative approach. It is also consistent with a recent variational Monte Carlo (VMC) study, another non-perturbative approach~\cite{WatanabeVMC}.

\begin{figure}
	\includegraphics[width=9cm]{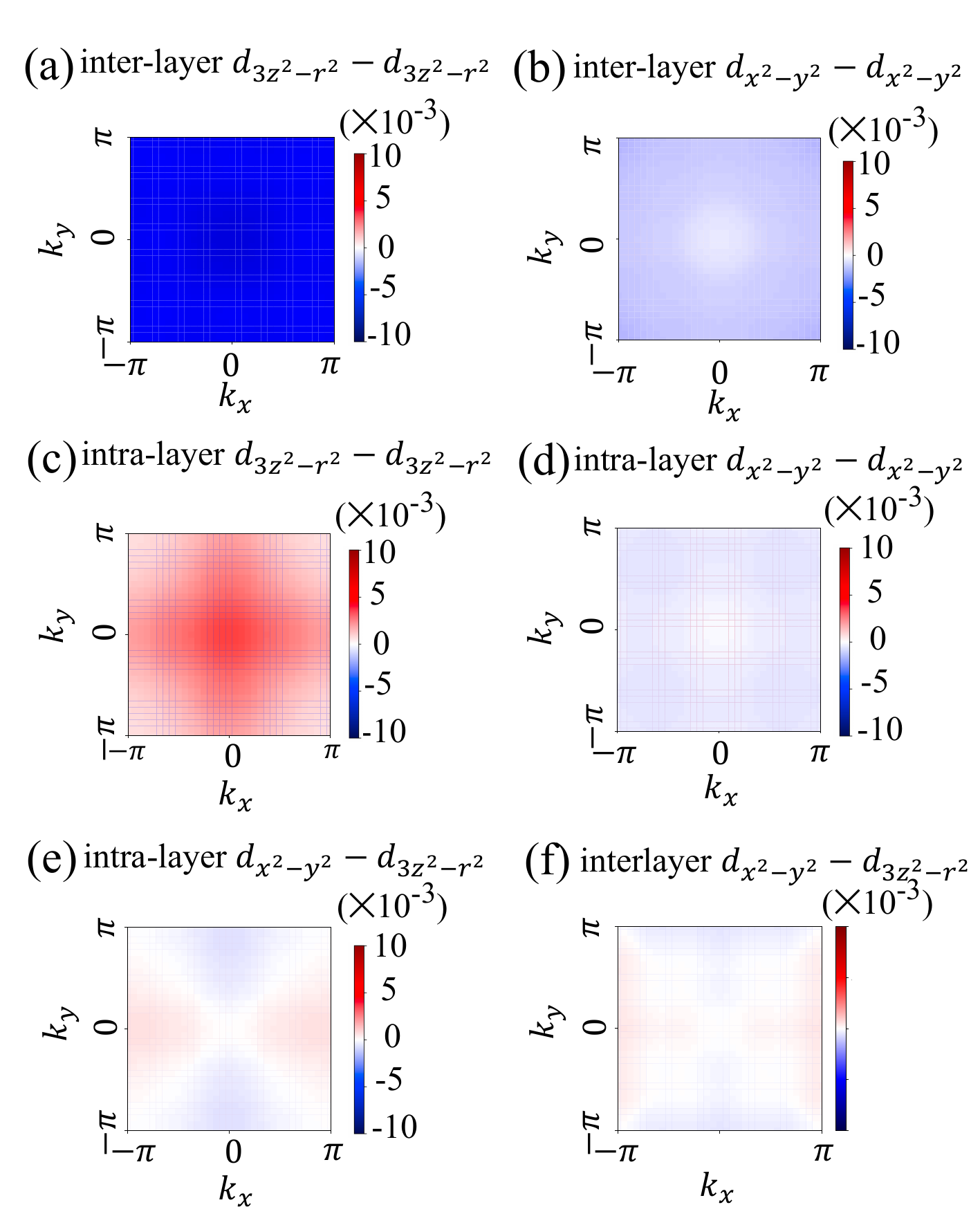}
	\caption{
	The gap functions in the orbital representation for the case of DFT$+U$ ($n=1.5$) at the $k_z=0$ cross section are presented on the two-dimensional Brillouin zone. Panels (a)-(f) display inter-layer diagonal $d_{3z^2-r^2}$, inter-layer diagonal $d_{x^2-y^2}$, intra-layer diagonal $d_{3z^2-r^2}$,
	intra-layer diagonal $d_{x^2-y^2}$, intra-layer off-diagonal $d_{x^2-y^2}$-$d_{3z^2-r^2} $, and interlayer off-diagonal $d_{x^2-y^2}$-$d_{3z^2-r^2}$ components, respectively.
	}
	\label{fig7}
\end{figure}

To understand the gap function in the band representation more clearly and to show more explicitly that superconductivity is dominated by the $d_{3z^2-r^2}$ orbitals, in Fig.~\ref{fig7}, we present the gap function in the orbital representation for the case of DFT$+U$. The $d_{3z^2-r^2}$ intraorbital interlayer pairing  component is by far the largest and nearly momentum independent, while the gap functions that involve the $d_{x^2-y^2}$ orbitals are small. The nearly-momentum-independent $d_{3z^2-r^2}$ gap function in the orbital representation is consistent with the fact that the gap function in the band representation is large wherever there is $d_{3z^2-r^2}$ orbital character in the bands. These results indicate that superconductivity is dominated by $d_{3z^2-r^2}$ orbital interlayer pairing that takes place mainly within a unit cell, as in the bilayer Hubbard model with large $U$ and $t_\perp$ near half filling~\cite{KA}.  In other words, it can be considered that in La$_3$Ni$_2$O$_7$, the bilayer Hubbard model formed by the $d_{3z^2-r^2}$ orbitals is {\it embedded} in a multiorbital system. 
The present calculation is consistent with a recent dynamical cluster approximation (DCA) study~\cite{MaierDagotto}, which is another non-perturbative approach.

Both the robustness of the $s\pm$-wave gap function and the nearly momentum-independent interlayer $d_{3z^2-r^2}$ gap function in the orbital representation are consequences of the finite energy spin fluctuations being the pairing glue, as will be discussed in detail in Sec.~\ref{fesf}.

Some comments are in order here. First, we stress that even when the $\gamma$ pocket is absent, namely, when the top of the $d_{3z^2-r^2}$ bonding band sinks below the Fermi level, the $d_{3z^2-r^2}$ orbitals are not half filled due to the $d_{3z^2-r^2}$-$d_{x^2-y^2}$ hybridization, and the $d_{3z^2-r^2}$ orbital electrons cannot be treated as localized spins.  In fact, the very existence of the split inner (so-called $\alpha$) and outer ($\beta$) Fermi surfaces observed experimentally~\cite{PengARPES,QKXue2502.17831,QKXue2502.17831}, which arises from the combination of interorbital hybridization and the large interlayer hopping $t_\perp$ of the $d_{3z^2-r^2}$ orbitals, indicates the itinerancy of the $d_{3z^2-r^2}$ electrons; if it were not for the interorbital hybridization, the two Fermi surfaces originating purely from the $d_{x^2-y^2}$ orbitals would be nearly degenerate~\cite{sakakibarala327}.

Second, in some previous theoretical studies~\cite{DMRG-Kaneko,DMRG-Kakoi,cDMFT-PRB109-165154,DMRG-ChinPhysLett40-127401,DMRG-PRL132-036502,DMRG-2311.12769}, the anomalous {\it Green function} of the $d_{x^2-y^2}$ orbitals (symbolically $\langle c^{\dagger} c^{\dagger}\rangle$) or its correlation functions ( $\langle c^{\dagger} c^{\dagger} c c\rangle$) are found to be comparable to or even larger than those of the $d_{3z^2-r^2}$ orbitals, in some cases even in the absence of Hund's coupling. In the present calculation (as well as in the non-perturbative approaches in Refs.~\cite{RyeeDMFT,MaierDagotto}), the gap function corresponds to the anomalous {\it self energy}, which is different from the anomalous Green function $\langle c^{\dagger} c^{\dagger}\rangle$ in that the pairing interaction $V_{\rm pair}$ is explicitly included as $V_{\rm pair}\langle c^{\dagger} c^{\dagger}\rangle$. The qualitative differences in these quantities indicate that the pairing interaction itself essentially arises only in the $d_{3z^2-r^2}$ orbital channel, while the phase coherence is passed on to the $d_{x^2-y^2}$ orbitals through the strong interorbital hybridization.

\begin{figure}
	\includegraphics[width=8cm]{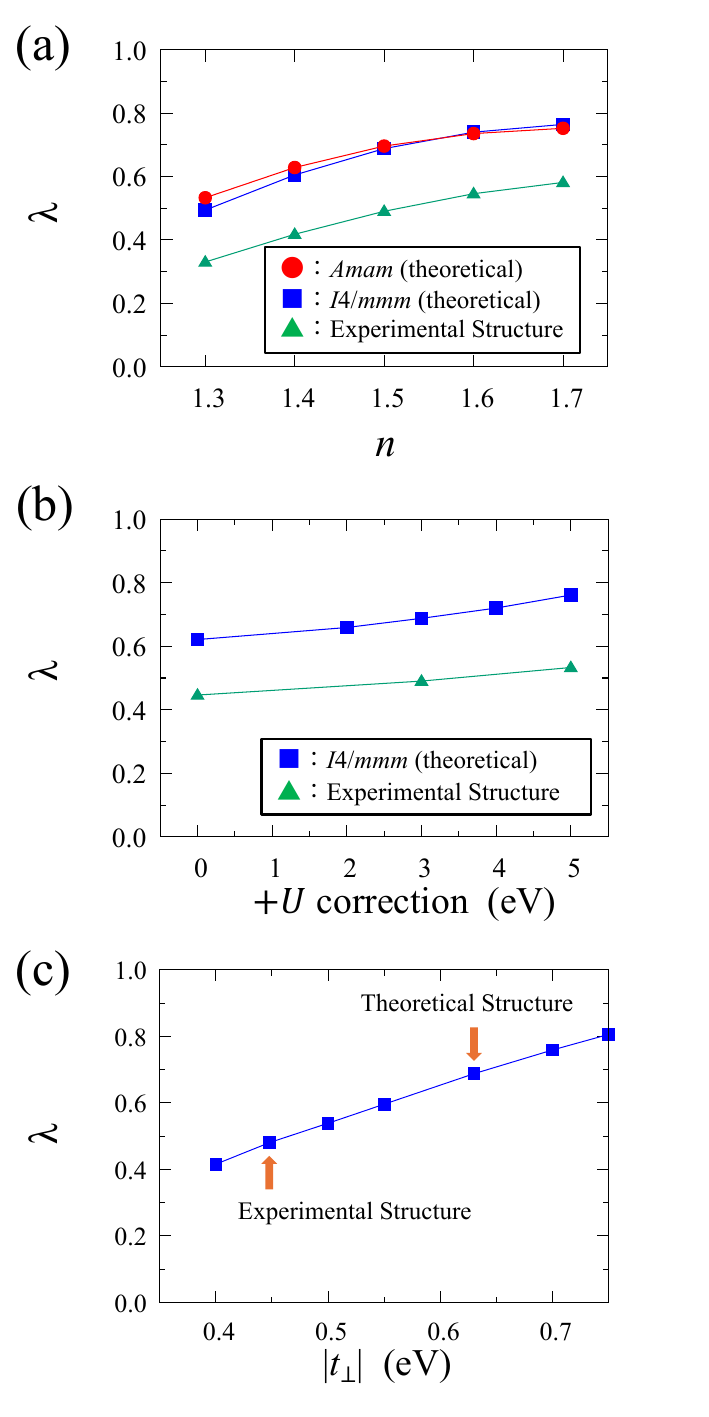}
	\caption{
	(a) The eigenvalues $\lambda$ of the linearized Eliashberg equation against the 
	band filling (the number of electrons per Ni atom) for the $Amam$(circles) and $I4/mmm$(squares) cases, 
	where $n=1.5$ corresponds to the stoichiometric case.
	$\lambda$ obtained for crystal structure experimentally determined in Ref.~\cite{QKXueThin} are also plotted (triangles).
	(b) $\lambda$ against $U$ for DFT$+U$ calculations at $n=1.5$ are shown.
	(c) $\lambda$ plotted against the interlayer hopping $t_\perp$ varied by hand. Here, all the other parameters are fixed at the values obtained from the theoretically optimized structure.
	}
	\label{fig8}
\end{figure}

\subsubsection{Eigenvalue of the linearized Eliashberg equation}

We now turn to the eigenvalue of the linearized Eliashberg equation $\lambda$ at $T=0.01$eV, which we adopt as a measure for $T_c$. In Fig.~\ref{fig8}(a), we plot $\lambda$ against the band filling $n$, where $n=1.5$ corresponds to the stoichiometric case. The results for $Amam$ ($\alpha=176^\circ$) and $I4/mmm$ cases ($\alpha=180^\circ$) are almost identical, which means that the deviation of $\alpha$ less than 5 degree have small impact on superconductivity within the present FLEX formalism. This  is due to the consideration of both the self energy and the full momentum and frequency dependencies of the Green function, which smears out the effect of the details in the band structure and the Fermi surface.
We also plot  in Fig.~\ref{fig8}(b)  the dependence of $\lambda$ on the $+U$ correction. The $+U$ correction dependence is not negligible but the variation is moderate.

In all the cases considered here, $\lambda$ monotonically increases as the band filling is increased. This is because the band filling of the $d_{3z^2-r^2}$ orbital, $n_{3z^2-r^2}$, approaches half filling~\cite{sakakibarala327}. In particular, regarding the relation between $\lambda$ (or $T_c$) and the presence/absence of the $\gamma$ pocket, $\lambda$ increases monotonically upon increasing the band filling $n$ (e.g. from $n=1.3$ to 1.5 or from $n=1.5$ to 1.6), without anything special happening at $n=1.5$, where the $\gamma$ pocket vanishes in the $+U = 3$ eV case. Also, $\lambda$ {\it decreases} from $n=1.5$ to $n=1.3$, where a large $\gamma$ pocket {\it appears} in the case of the experimental structure. Hence, ``the better the Fermi surface nesting, the better for superconductivity'' does not hold as far as the present formalism for the bilayer system is concerned. Once again, this is due to the fact that the finite energy spin fluctuations are playing an important role in enhancing superconductivity~\cite{Nakata}, as will be discussed in detail in Sec.~\ref{fesf}.

In contrast to the $+U$ correction dependence, or the comparison between $Amam$ and $I4/mmm$ cases, the difference of $\lambda$ between experimental and theoretical structure is significant. 
The suppression of $\lambda$ for the experimental structure cannot be traced back to the difference in $n_{3z^2-r^2}$; we have checked that $n_{3z^2-r^2}$ barely changes between theoretical and experimental structures. In fact, $\lambda$ monotonically decreases if we reduce $|t_\perp|$ by hand, as shown in Fig.~\ref{fig8}(c). The origin of this behavior can once again be explained from the  effect of the finite energy spin fluctuations, as will be discussed in Sec.~\ref{fesf}.

Since $\lambda=0.50$~\cite{mrpa2} is obtained for the cuprates with relatively low $T_c$ (La$_2$CuO$_4$, $T_c\simeq 40$K) in the same calculation scheme, the $\lambda$ values obtained here for the experimental crystal structure may explain the reduced $T_c$ in thin films (see Fig.~\ref{figX} in the Appendix). Even if this is the case, the reason why the crystal structures determined theoretically and experimentally differ so much remains as an interesting puzzle. 

On the other hand, $\lambda$ values obtained for theoretical structures do not differ much from what we obtained for the pressurized bulk~\cite{sakakibarala327}. The reason for this is because, although $|t_\perp|$ is somewhat smaller in thin films, $n_{3z^2-r^2}$ is larger (0.95 for the SLAO substrate vs. $\sim 0.9$ for the pressurized bulk), which compensates the effect of the reduction of $|t_\perp|$. Regarding this failure of the theoretical structure to explain the $T_c$ reduction in thin films, let us note the following. The $Amam$ and $I4/mmm$ cases give essentially the same values of $\lambda$ within the present scheme, but this might be because we are not taking into account the effect of the competing diagonal order such as density waves. We still should not rule out the possibility that the symmetry of the actual material is somewhat lowered from $I4/mmm$, where (short range) diagonal order that competes with superconductivity might be present to reduce $T_c$. This problem is beyond the scope of the present study, and also remains as an interesting future study.

\subsubsection{Finite energy spin fluctuations}
\label{fesf}

\begin{figure}
	\includegraphics[width=8.5cm]{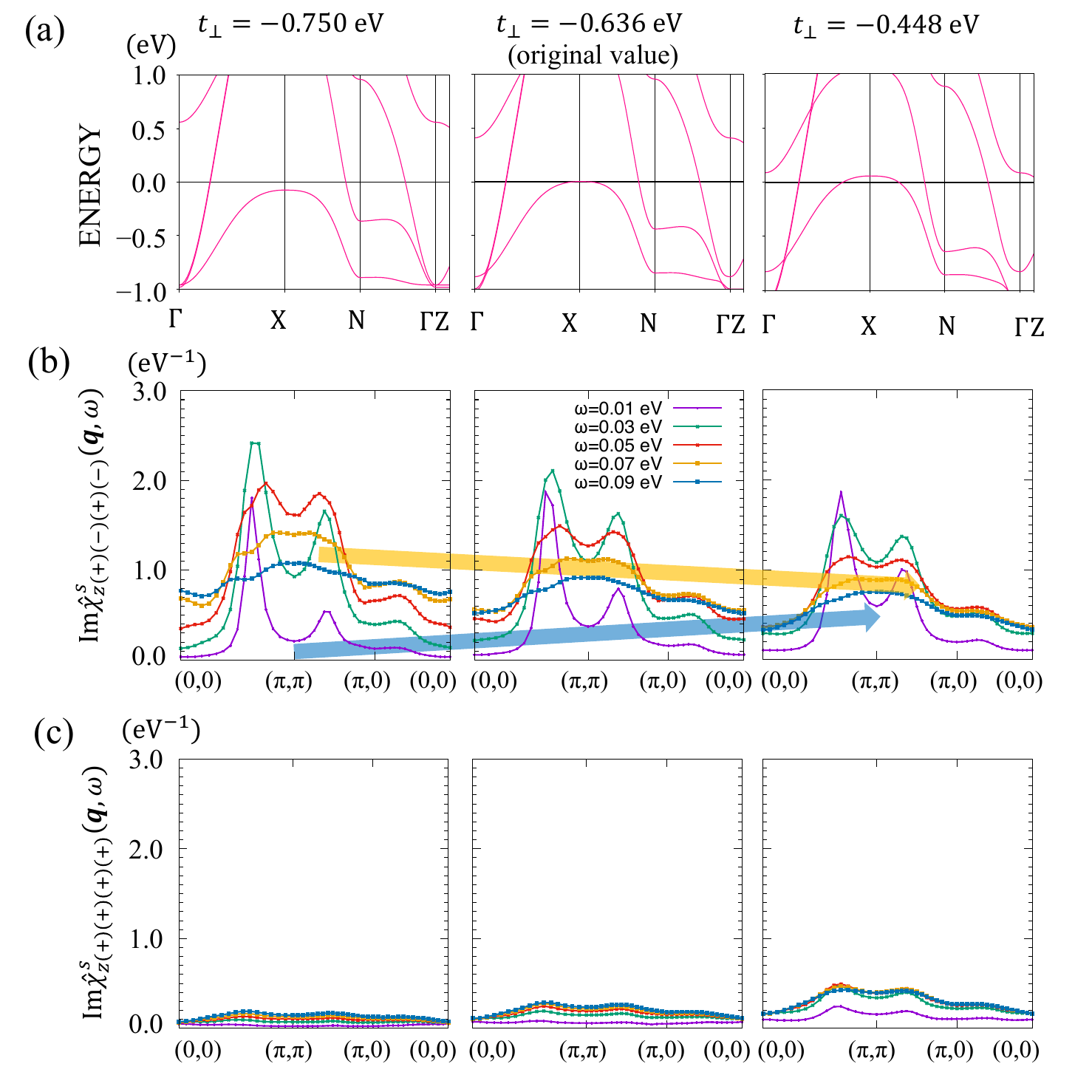}
	\caption{Left, center and right panels corresponds to $|t_{\perp}|=$ 0.750, 0.638, and 0.448 eV, respectively.  Panels (a) : the bare band structure. Panels (b) : the imaginary part of the $d_{3z^2-r^2}$ interband dynamical spin susceptibility $\hat \chi _{z(+)z(-)z(+)z(-)}^{s}({\bm q},\omega)$ plotted along the high symmetry line within the Brillouin zone for different values of $\omega$. Panels (c) : plots similar to (b) for the intraband component $\hat \chi _{z(+)z(+)z(+)z(+)}^{s}({\bm q},\omega)$. We note that the energy scale of the spin fluctuations do not correspond quantitatively to the energy scale of the bare bands because the former is governed by the energy scale of the renormalized bands.
	}
	\label{fig9}
\end{figure}

We have mentioned in the above that finite energy spin fluctuations play an important role in $s\pm$-wave pairing. To show this explicitly, in Fig.~\ref{fig9}, we present the imaginary part of the dynamical spin susceptibility $\hat \chi^{s}({\bm q},\omega)$ obtained from the FLEX results through analytical continuation using Pad\'e approximation. We consider three cases for $t_\perp$ (considered in Fig.~\ref{fig8}(c) in the previous section), and present the interband and intraband $d_{3z^2-r^2}$ orbital components of the spin susceptibility as was done in Refs.~\cite{RyeeDMFT, RyeeRPA}. Namely, we first obtain the spin susceptibility $\hat \chi^{s}$ in orbital basis and then transform it to that of the bonding-antibonding basis.
The matrix elements of $A$ in Eq.~(\ref{unimat}) gives the following unitary transformation,
\begin{equation}
\hat \chi _{ijkl}^{s}({\bm q},i\omega_n)=\sum _{a,b,c,d} A_{ia} A_{jb}A_{kc} A_{ld} \hat \chi_{abcd}^s ({\bm q},i\omega_n),\label{chisiomega}
\end{equation}
where $i,j,k,l \in \{ x(+),  z(+), x(-) , z(-)\}$ and 
$a,b,c,d \in \{ x({\rm U}) ,z({\rm U}),x({\rm L}),z({\rm L})\}$.
Then we continuate $\hat \chi _{ijkl}^{s}({\bm q},i\omega_n)$ to $\hat \chi _{ijkl}^{s}({\bm q},\omega)$.
We call off-diagonal components
\begin{equation}
\hat \chi _{z(+)z(-)z(+)z(-)}^{s}({\bm q},\omega)
\end{equation}
between bonding-antibonding basis as {\it interband} components for the sake of convenience.
The interband component of the spin fluctuations acts as a pairing glue for the $s\pm$ pairing, where the gap function changes its sign between the bonding and antibonding bands. 

It can be seen that the low energy interband spin fluctuations with $\omega$ equal to or smaller than $0.03$ eV are strongly momentum dependent, reflecting the nesting of the Fermi surface. Namely, the peak close to $(\pi/2,\pi/2)$ originates from the nesting between $\alpha$ and $\beta$ Fermi surfaces~\cite{Luo}, the peak around $(\pi/2,0)$ from the nesting between $\beta$ and $\gamma$, and the low energy spin fluctuations around $(\pi,\pi)$ originates from the interaction between $\gamma$ around $(\pi,\pi)$ and the antibonding $d_{3z^2-r^2}$ band around $(0,0)$. By contrast, the relatively high energy interband spin fluctuations with $\omega$ larger than $\sim 0.05$ eV are less momentum dependent. This is because the interband interaction involving states away from the Fermi level with various combinations of bonding and antibonding wave vectors can give rise to spin fluctuations with similar $\omega$. 
The fact that the orbital representation of the gap function for the interlayer $d_{3z^2-r^2}$ pairing is  nearly momentum independent (i.e. intra-unit-cell pairing) suggests that the present $s\pm$-wave pairing is mainly dominated by the relatively high energy spin fluctuations, similarly to the case of the bilayer Hubbard model studied in Ref.~\cite{Nakata}. 

The above view is also consistent with the correlation found among $t_\perp$, $\lambda$, and the spin fluctuations. Namely, as $|t_{\perp}|$ is reduced, the interband component of the low energy spin fluctuations around $(\pi,\pi)$ and $(\pi,\pi/2)$ gets enhanced. This is because the $\gamma$ Fermi surface  appears and also the antibonding $d_{3z^2-r^2}$ band around $(0,0)$ comes closer to the Fermi level, so that the Fermi surface nesting involving these bands becomes better. On the other hand, the relatively high energy spin fluctuations are suppressed by reducing $t_\perp$. Thus, the suppression of $\lambda$ upon reducing $|t_\perp|$ can be naturally understood if the high energy spin fluctuations, rather than the low energy ones, are playing a major role in the pairing. (It should also be noted that the {\it intraband} component of the finite energy spin fluctuations is somewhat enhanced when $|t_{\perp}|$ is small, which is also not favorable for $s\pm$ superconductivity, where the gap sign is maintained within each band.)

These calculation results of the dynamical spin susceptibility indicate that the present $s\pm$-wave pairing is dominated by finite energy spin fluctuations, not by the Fermi surface nesting in the narrow sense of the term. Superconductivity being insensitive to the presence/absence of the $\gamma$ Fermi surface is consistent with a recent VMC study~\cite{WatanabeVMC}.

Experimentally, the relevance of these finite-energy spin fluctuations is supported by recent inelastic neutron scattering spectra, suggesting strong magnetic excitations originating from the interlayer coupling~\cite{Xie_neutron}.

\section{Summary}

We have theoretically investigated ambient pressure superconductivity in thin films of La$_3$Ni$_2$O$_7$ using the combination of first-principles calculation and FLEX calculation.
As the in-plane lattice constant $a$ is reduced, the interlayer Ni-O-Ni bond angle approaches 180 degrees, qualitatively consistent with previous theoretical studies~\cite{BotanaThinPRB,2502.01624,85qv-ncxb}. Also, the number of electrons in the $d_{3z^2-r^2}$ orbital increases as $a$ is reduced due to the increase of the lattice constant $c$ and the inner apical oxygen height, and hence the increase of the level offset between the two orbitals.

For the SLAO substrate in particular, we have found that the deviation of the interlayer Ni-O-Ni bonding angle from 180$^\circ$ has small impact on superconductivity as far as the present FLEX approach is concerned. 
Although the size and the presence of the $\gamma$ pocket largely depends on the crystal structure adopted and/or the presence/absence of $U$ correction in the PBEsol+$U$ scheme, $s\pm$-wave pairing symmetry remains robust, and the gap function is always essentially the same as what we obtained for the pressured bulk system~\cite{sakakibarala327}. This is due to the fact that the pairing is mediated mainly by finite energy spin fluctuations~\cite{Nakata}, which have relatively small momentum dependence, resulting in a gap function of the interlayer $d_{3z^2-r^2}$ pairing that is nearly momentum independent, i.e., intra-unit-cell pairing in real space. The nearly-wave-number-independent gap function for the interlayer pairing of the $d_{3z^2-r^2}$ orbitals is essentially the same as that obtained for the simple bilayer Hubbard model~\cite{KA}, so the present results indicate that La$_3$Ni$_2$O$_7$ can be viewed as a system where the bilayer Hubbard formed by the $d_{3z^2-r^2}$ is embedded in a multiorbital system.
We stress that a formalism that takes into account the full momentum and frequency dependencies of the Green function and the pairing interaction is important for treating the effect of such finite energy spin fluctuations. The overall features of the gap function obtained by FLEX well agrees with non-perturbative approaches such as cDMFT~\cite{RyeeDMFT}, DCA~\cite{MaierDagotto}, and VMC~\cite{WatanabeVMC}.

The reason why $T_c$ is halved from that of the pressurized bulk cannot be understood by adopting the theoretically determined crystal structure, at least within the present formalism, although there may be other origins (e.g. competition with density waves) for the reduced $T_c$ that are not taken into account in the present scheme. On the other hand, the $T_c$ reduction in thin films can be understood by adopting the experimentally determined crystal structure of Ref.~\cite{QKXueThin}, for which $|t_\perp|$ is small. Our analysis shows that the reduction of $\lambda$ for the experimental structure can be traced back to the $t_\perp$ dependence of $T_c$ of the $s\pm$-wave superconductivity in the single orbital bilayer Hubbard model. Even if this is the case, the origin of the large discrepancy between experimental and theoretical crystal structures remains as an interesting future problem.

\section*{acknowledgments}
We thank Dr. Hiroshi Watanabe and Dr. Motoki Osada for fruitful discussions.
This work was supported by JST K Program Grant No. JPMJKP25Z3, JSPS KAKENHI Grant No. JP24K01333, JP25K08459 (H.S.), JP22K04907, JP25K00959, JP26K08179 (K.K), and JST-FOREST Program Grant No. JPMJFR212P (M.O.), JPMJFR246T (H.S.).
The computing resource is supported by the supercomputer system (system-B) in the Institute for Solid State Physics, the University of Tokyo, and the supercomputer of Academic Center for Computing and Media Studies (ACCMS), Kyoto University.

\appendix*

\section{Formalism of multi-orbital FLEX and comparison with experiments}

We briefly summarize the formalism of multi-orbital FLEX adopted in this study.
We consider a multi-orbital Hamiltonian given in the present form,
\begin{eqnarray}
H_{\mathrm{int}}=&&U\sum_{i,\mu}n_{i\mu\uparrow}n_{i\mu\downarrow}
+U'\sum_{i,\mu<\nu,\sigma}n_{i\mu\sigma}n_{i\nu\bar{\sigma}}\nonumber\\ 
&&+(U'-J)\sum_{i,\mu<\nu,\sigma}n_{i\mu\sigma}n_{i\nu\sigma}\nonumber\\
&&-J\sum_{i,\mu\neq\nu}c^{\dagger}_{i\mu\uparrow}c_{i\mu\downarrow}c^{\dagger}_{i\nu\downarrow}c_{i\nu\uparrow}\nonumber\\
&&+J'\sum_{i,\mu\neq\nu}c^{\dagger}_{i\mu\uparrow}c^{\dagger}_{i\mu\downarrow}c_{i\nu\downarrow}c_{i\nu\uparrow},
\end{eqnarray}
where $U$, $U'$, $J$, and $J'$ are the intraorbital repulsion, the interorbital repulsion, Hund's coupling, and the pair hopping, respectively. 
Here, $i$ denotes the sites, $\mu$, $\nu$ the orbitals, and $\sigma$ the spins.
$c_{i\mu\sigma}$ and $c^{\dagger}_{i\mu\sigma}$ are annihilation and creation operators of electrons, respectively. 
The number operator is defined as $n_{i\mu\sigma}=c^{\dagger}_{i\mu\sigma}c_{i\mu\sigma}$.

In the FLEX approximation, the self energy is calculated by taking into account the bubble and ladder type diagrams that comprise the irreducible susceptibility 
\begin{equation}
\chi^{0}_{l_{1}l_{2}l_{3}l_{4}}(q)=-\frac{T}{N}\sum_{k}G_{l_{1}l_{3}}(k+q)G_{l_{4}l_{2}}(k),
\end{equation}
which is calculated from the renormalized Green function $G_{l_il_j}(k)$, where $k,q$ stands for both the wave vector and the Matsubara frequency, and $l_{i}$ denotes the orbitals. 
The Dyson equation is numerically solved to obtain the renormalized Green function in a self-consistent manner.
Using the converged Green function and the spin and charge susceptibilities,  
\begin{equation}
\hat{\chi}^{s}(q)=\frac{\hat{\chi}^{0}(q)}{1-\hat{S}\hat{\chi}^{0}(q)},
\end{equation}
\begin{equation}
\hat{\chi}^{c}(q)=\frac{\hat{\chi}^{0}(q)}{1+\hat{C}\hat{\chi}^{0}(q)},
\end{equation}
respectively, with the interaction matrices
\begin{equation}
S_{l_{1}l_{2}l_{3}l_{4}}=\left\{\begin{array}{cc}
U, & l_{1}=l_{2}=l_{3}=l_{4},\\
U', & l_{1}=l_{3}\neq l_{2}=l_{4},\\
J, & l_{1}=l_{4}\neq l_{2}=l_{3},\\
J', & l_{1}=l_{2}\neq l_{3}=l_{4},
\end{array}\right.
\end{equation}
\begin{equation}
C_{l_{1}l_{2}l_{3}l_{4}}=\left\{\begin{array}{cc}
U, & l_{1}=l_{2}=l_{3}=l_{4},\\
-U'+2J, & l_{1}=l_{3}\neq l_{2}=l_{4},\\
2U'-J, & l_{1}=l_{4}\neq l_{2}=l_{3},\\
J', & l_{1}=l_{2}\neq l_{3}=l_{4},
\end{array}\right.
\end{equation}
are plugged into the linearized Eliashberg equation
\begin{eqnarray}
  &&\lambda\Delta_{\mu\nu}(k)\nonumber\\
  &&=-\frac{T}{N}\sum_{q,m_1,m_2,m_3,m_4}V_{{\rm pair},\mu m_{1}m_{4}\nu}(q)G_{m_{1}m_{2}}(k-q)\nonumber\\
  &&\times \Delta_{m_{2}m_{3}}(k-q)G_{m_{4}m_{3}}(q-k)
\end{eqnarray}
to obtain eigenvalue $\lambda$ and eigenfunction $\Delta(k)$, where the effective spin-singlet pairing interaction is
\begin{equation}
V_{\rm pair}(q)=\frac{3}{2}\hat{S}\hat{\chi}^{s}(q)\hat{S}-\frac{1}{2}\hat{C}\hat{\chi}^{c}(q)\hat{C}+\frac{3}{4}\hat{S}+\frac{1}{4}\hat{C}.
\end{equation}

\begin{figure}
	\includegraphics[width=9cm]{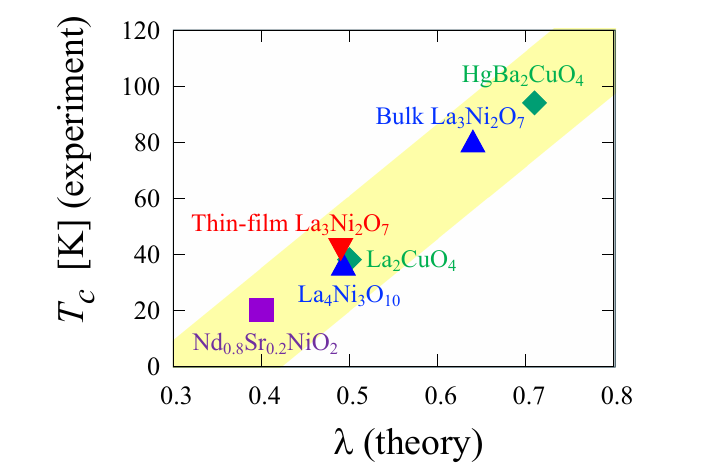}
	\caption{
	$T_c$'s are plotted against the eigenvalues of Eliashberg equation $\lambda$ at a fixed temperature, $T=0.01$ eV.
	Symbols of diamond, square, upward triangle indicate
	cuprates, infinite-layer nickelates, bulk Ruddlesden-Popper type nickelates, respectively.
	The downward triangle indicates $\lambda$ of thin film La$_3$Ni$_2$O$_7$, obtained by using the experimental structure provided by Yue {\it et al}.~\cite{QKXueThin}
	(see the main text for details).
	Yellow hatched region is a guide to the eyes.
	}
	\label{figX}
\end{figure}

To demonstrate the validity of the present formalism, we plot in Fig.~\ref{figX} the calculated $\lambda$ vs. the experimentally observed $T_c$ for some unconventional superconductors.
Regarding the interaction parameters $U, U', J,$ and $J'$, we adopted values calculated within the constrained random phase approximation (cRPA)~\cite{cRPA}, except for the cuprates. For the cuprates, these parameters were determined using the model-mapped RPA (mRPA)~\cite{mrpa}, which yields values comparable to those from cRPA, as discussed in Ref.~\cite{mrpa2}. In the specific case of $\text{La}_4\text{Ni}_3\text{O}_{10}$, we used the parameters calculated for $\text{La}_3\text{Ni}_2\text{O}_7$ via cRPA \cite{Werner}.
As for the experimental data, $T_c$'s for $\text{La}_2\text{CuO}_4$ \cite{LSCO}, $\text{HgBa}_2\text{CuO}_4$ \cite{HBCO}, $\text{Nd}_{0.8}\text{Sr}_{0.2}\text{NiO}_2$ \cite{Hwang}, bulk $\text{La}_3\text{Ni}_2\text{O}_7$ \cite{MWang}, $\text{La}_4\text{Ni}_3\text{O}_{10}$ \cite{NagataLa4310}, and thin-film $\text{La}_3\text{Ni}_2\text{O}_7$ \cite{Ko2025} were sourced from their respective literatures.
The theoretical values of $\lambda$ for $\text{La}_2\text{CuO}_4$ and $\text{HgBa}_2\text{CuO}_4$ were taken from Ref. \cite{mrpa2}, while that of $\text{Nd}_{0.8}\text{Sr}_{0.2}\text{NiO}_2$ was obtained from Ref. \cite{SakakibaraNi}. The $\lambda$ values for bulk $\text{La}_3\text{Ni}_2\text{O}_7$ and $\text{La}_4\text{Ni}_3\text{O}_{10}$ have been updated from previous reports~\cite{sakakibarala327,sakakibara4310}; these were recalculated using DFT+$U$ with $U=3$ eV to ensure consistency with the present study.
The crystal structures used in each calculation were obtained either through theoretical optimization (for bulk $\text{La}_3\text{Ni}_2\text{O}_7$ and $\text{La}_4\text{Ni}_3\text{O}_{10}$) or from experimental observations reported in the literature for $\text{La}_2\text{CuO}_4$ \cite{Jorgensen}, $\text{HgBa}_2\text{CuO}_4$ \cite{Wagner}, $\text{Nd}_{0.8}\text{Sr}_{0.2}\text{NiO}_2$ \cite{HaywardNd}, and thin-film $\text{La}_3\text{Ni}_2\text{O}_7$ \cite{QKXueThin}.

\bibliography{thin}

\begin{thebibliography}{86}
\bibitem [{kco()}]{kconv}%
  \BibitemOpen
  \href@noop {} {}\bibinfo {note} {We have checked more dense ${\bm k}$-mesh up
  to a 12 $\times$ 12 $\times$ 12 ${\bm k}$-mesh. However, there is no
  significant differences with the 8 $\times$ 8 $\times$ 8 ${\bm k}$-mesh case,
  e.g., the difference of the lattice constant $c$ is about 0.01 \AA~~between
  the two cases. Therefore, we present the results of the 8 $\times$ 8 $\times$
  8 ${\bm k}$-mesh case in the main text.}\BibitemShut {Stop}%
\end{thebibliography}

\end{document}